# Injection locking at zero field in two free layer spin-valves

Mario Carpentieri,[1] Takahiro Moriyama,[2] Bruno Azzerboni,[3] Giovanni Finocchio[3]

[1]Department of Ingegneria Elettrica e dell'Informazione, Politecnico of Bari, via E. Orabona 4, I-70125 Bari, Italy.

[2]Institute for Chemical Research, Kyoto University, Uji, Kyoto, 611-0011, Japan

[3]Department of Electronic Engineering, Industrial Chemistry and Engineering. University of Messina, C.da di Dio, I-98166, Messina, Italy.

**Abstract**: This paper predicts the possibility to achieve synchronization (via injection locking to a microwave current) of spin-transfer torque oscillators based on hybrid spin-valves composed by two free layers and two perpendicular polarizers at zero bias field. The locking regions are attained for microwave frequency near $0.5f_0$, $f_0$, and $2f_0$ where $f_0$ is the input oscillator frequency. Those properties make this system promising for applications, such as high-speed frequency dividers and multipliers, and phase-locked-loop demodulators.

Persistent magnetization oscillations driven by a spin-polarized current has been extensively studied in spin-valves,[1,2] point-contact geometries,[3] and magnetic tunnel junctions.[4,5] Experiments indicate a very rich dynamical behavior for spin-transfer-torque oscillators (STOs) due to the excitation of uniform and non-uniform modes (such as vortex,[6] vortex-antivortex pair[7]). One of the key issues to put STOs into practical applications is to achieve the excitation of persistent dynamics at zero bias field in order to reduce the size and the power dissipation of the oscillator.[8] To date, the larger frequency (near 6 GHz) at zero bias field has been experimentally measured in hybrid spin valves composed of two coupled in-plane free layers and two out-of-plane polarizers in which the magnetization precession in the two free layers is phase locked and the GMR signal oscillates at twice the frequency of the mode excited in each free layer.[9] This hybrid spin-valve is promising from a technological point of view, showing several advantages over other STO solutions, such as wide current range of dynamics, high oscillation frequency at reduced linewidth and, in addition, a larger output power may be achieved if the two free layers are separated from a tunnel barrier instead of a normal metal spacer.

The key result of this work is the numerical prediction of zero field injection locking phenomena (to date it has been observed only in presence of bias fields[10,11,12,13]) driven by microwave currents in the same category of hybrid spin-valves studied in Ref.[9]. The locking regions are achieved for external microwave frequencies $f_{AC}$ near $f_0$ (first harmonic), $2f_0$ (second harmonic) and also at $0.5 f_0$ (half harmonic), where $f_0$ is the oscillator frequency. Those hybrid spin-valves can be used as frequency dividers, frequency multipliers and as a generalization of the conventional phase locking loop (PLL) devices, where the locking at fractional harmonic is used in the synchronization transient to increase the locking speed, while the *n*-integer harmonic synchronization is used in the locked regime to make simpler the control electric circuit.[14]

The active part of the spin-valves under investigation is composed by two in-plane free layers of cobalt and two perpendicular polarizers all separated by a copper spacer [CoNi (8) polarizer#1 / Cu (2) / Co (4) free layer #1 / Cu (4) / Co (4) free layer #2/ Cu (2) / CoNi or CoPt (8) polarizer #2] (the numbers in parentheses are the thicknesses in nm) as displayed in Fig. 1. To achieve the perpendicular polarizer, we consider the same material combination Co/Pt and Co/Ni used in the experiment of Ref.[9]. We studied different cross sections: i.e. elliptical (170 x 130 $nm^2$) and circular (diameter of 140 nm) and for different perpendicular polarizers: i.e. symmetric (both are CoNi) and asymmetric (CoNi is on the top and CoPt on the bottom).

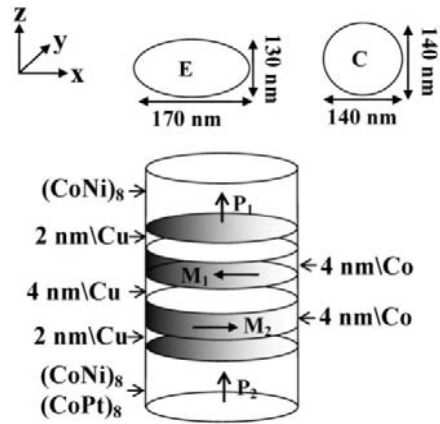

FIG. 1 Sketch of the device geometry. $P_1$ and $P_2$ indicate the directions of out-of-plane polarizers, while $M_1$ and $M_2$ are the in-plane magnetization of the free layers. The letters "E" and "C" specify the two different cross-sections.

The magnetization dynamics is modeled by using the Landau–Lifshitz–Gilbert-Slonczweski (LLGS) equation[15]. For the Co/Cu/Co tri-layer, we consider that the two free layers are coupled by the spin-transfer torque and the magnetostatic field and the spin-torque from each polarizer is acting on the neighboring Co layer (more computational details can be found in [9, 16]). Concerning the magnetoresistance signal, the contributions from the junctions with the two perpendicular polarizers are also included. The spin torques on the top and bottom free layer are given by:

$$\begin{cases} \mathbf{T}(\mathbf{m_t},\mathbf{m_b}) = \dfrac{g|\mu_B|j}{e\gamma_0 dM_s^2}\mathbf{m_t} \times \left[\varepsilon(\mathbf{m_t},\mathbf{m_b})(\mathbf{m_t}\times\mathbf{m_b}) - \varepsilon_{pt}(\mathbf{m_t}\times\mathbf{m_p})\right] \\ \mathbf{T}(\mathbf{m_b},\mathbf{m_t}) = -\dfrac{g|\mu_B|j}{e\gamma_0 dM_s^2}\mathbf{m_b} \times \left[\varepsilon(\mathbf{m_b},\mathbf{m_t})(\mathbf{m_b}\times\mathbf{m_t}) - \varepsilon_{pb}(\mathbf{m_b}\times\mathbf{m_p})\right] \end{cases} \quad (1)$$

where $\mathbf{m_t}$ and $\mathbf{m_b}$ are the magnetization unit vectors for the top and bottom Co-layers respectively, $g$ is the gyromagnetic splitting factor, $\mu_B$ is the Bohr magneton, $j = j_{DC} + j_M \sin(2\pi f_{AC} t)$ is the current density including direct and microwave terms, $d$ is the thickness of the Co layer, $e$ is the electron charge, $\mathbf{m_p}$ is a unit vector in the direction of the out-of-plane polarizer, we assume that the magnetizations of the two perpendicular polarizer layers remain pinned along the $+z$ direction (see Fig. 1). $\varepsilon(\mathbf{m_t},\mathbf{m_b}) = \eta/(1+B\mathbf{m_b}\bullet\mathbf{m_t})$ is the polarization function computed by Slonczweski for symmetric spin valves (rewritten in a simpler form), $\varepsilon_{pt}$ and $\varepsilon_{pb}$ are the spin current polarizations for the Co/Pt and the Co/Ni respectively. We employ the values $k_u$=1.26 x $10^4$ J/m$^3$ for uniaxial magneto-crystalline anisotropy along the easy axis, saturation magnetization $M_S$ =1.8 T for the Co layer, 0.63 T for the Co/Pt multilayer, and 0.82 T for the Co/Ni multilayer, and values for the perpendicular anisotropy constants of 3.3 x $10^5$ J/m$^3$ for the Co/Ni and 20 x $10^5$ J/m$^3$ for the Co/Pt, $\eta$=0.36 and $B$=0.66 are the spin current polarization and the asymmetric parameter for the Co layer. In our systematic study, we consider for the Co layers exchange constant $A$=2.0, 2.4 and 2.8x$10^{-11}$ J/m and for the Gilbert damping $\alpha_G$=0.01, 0.02, 0.04. Since the numerical results with those different parameters are found to be qualitatively the same, in the rest of the paper we would like to focus on the detailed discussion for the results with $A$=2.0x$10^{-11}$ J/m and $\alpha_G$=0.02.

The first finding is the possibility to control the slope of the frequency vs current curve by adjusting the trade-off between the spin-torque from the in-plane Co layers and the two perpendicular polarizers (not shown). Here, we point the attention to the results of injection locking achieved when the oscillator frequency vs current exhibits blue shift (the locking regions in the blue shift case are larger than the ones in the red shift behavior). In particular, the data in the rest of the letter are with $\varepsilon_{pt}$=0.2 and $\varepsilon_{pb}$=0.4, similar qualitative results are achieved for $\varepsilon_{pt}$=0.18 and 0.22 and for $\varepsilon_{pb}$=0.35 and 0.45.[18] We underline that exists a source of asymmetry also in the symmetric devices related to the torque between the two in-plane free layers.

The first step is the characterization of the free running data, $j_M$=0 A/cm$^2$. Fig. 2 summarizes the results of the oscillation frequency as a function of the current density $j_{DC}$ on the basis of micromagnetic simulations for AE (asymmetrical polarizers and elliptical shape), SE (symmetrical polarizers and elliptical shape), AC (asymmetrical polarizers and circle shape) and SC (symmetrical polarizers and circle shape) respectively. The dynamical behaviour observed is qualitatively similar to the one described in our previous paper (Ref. 9). In particular, the magnetization precession is characterized by a phase locking between the magnetization precessions in the two free layers which oscillate at the same frequency, resulting GMR oscillation at twice that frequency. At high current the magnetization oscillation is not achieved and a vortex state is nucleated in both free layers.[9, 19] The minimal critical current $J_C$=1.1 x $10^7$ A/cm$^2$ is achieved for the SC.

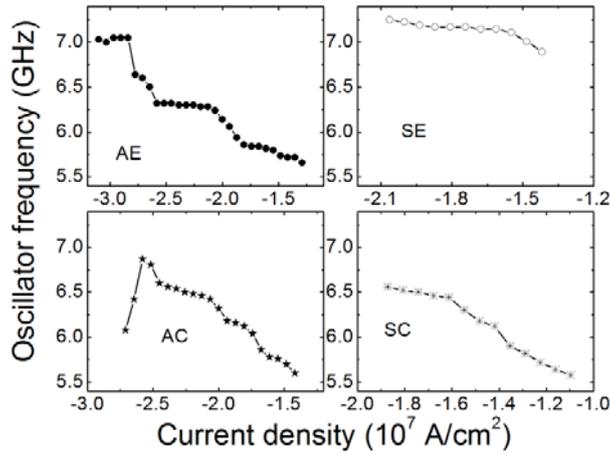

FIG. 2 Free running data (oscillation frequency as a function of current density) computed by means of the micromagnetic simulations. AE, SE, AC, and SC indicate Asymmetrical polarizers and Elliptical shape, Symmetrical polarizers and Elliptical shape, Asymmetrical polarizers and Circular shape, and Symmetrical polarizers and Circular shape respectively.

Starting from the free running properties, the synchronization mechanism (non-autonomous dynamics) is systematically studied. The injection locking near the oscillation frequency (GMR-signal) has been already observed in presence of external fields for both microwave currents[10] and fields.[20] Those results have been described with success on the basis of micromagnetic simulations[21] and analytical theories.[22, 23, 24]

Our numerical experiments point out the presence of the locking phenomenon at the first harmonic $f_0$ in the whole range of current where magnetization dynamics is observed and for all the four different device configurations. Those results are not discussed in detail being similar to what has been already observed in literature. In particular, the predict locking regions are of the same order of the ones achieved for different devices such as point-contact geometries and magnetic tunnel junctions biased with a bias field (for example for the SC the locking region is 180 MHz at $J_M$=3.9 x $10^6$ A/cm$^2$).[10, 11] However, we stress the fact that the main difference in our results with respect to the ones already published, which is also an important technological advantage, is the locking or synchronization achieved at zero bias field.

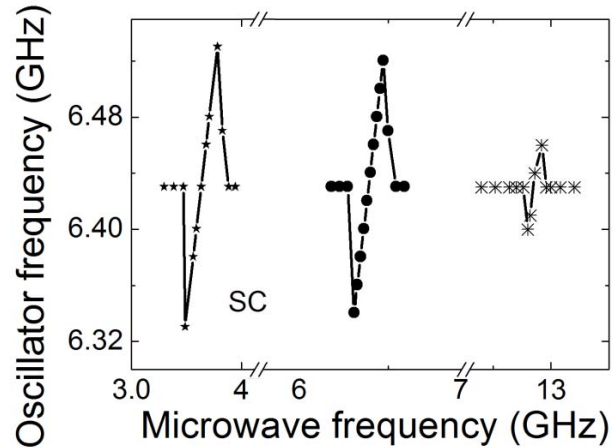

FIG. 3 Dependence of the oscillator frequency on the microwave frequency (SC spin valve) for $J_{DC}$=-1.6 x $10^7$ A/cm$^2$ and $J_M$=3.9 x $10^6$ A/cm$^2$.

The key result of our numerical study is the achievement of the synchronization also at half and two times the input oscillator frequency $f_0$, i.e. half-harmonic (0.5$f_0$) and 2-harmonic (2$f_0$) synchronization respectively. In some cases, this locking behavior is achieved for the same bias current as shown for the example in Fig. 3, where the oscillator frequency as function of the microwave frequency for the SC spin-valve achieved at $J_{DC}$=-1.6 x $10^7$ A/cm$^2$ and $J_M$=3.9 x $10^6$ A/cm$^2$ is displayed. This property is at the basis for the design of spintronic frequency dividers and multipliers.

A systematic study of the locking shows that this behavior is general and achieved for all the four different geometries and in a wide range of bias currents. Fig. 4(a) and (b) summarize the locking bandwidth as function of the amplitude of the microwave current (Arnold Tongue) for the four devices and different $J_{DC}$. In Fig. 4(a) the oscillator frequencies of $f_0$=5.82 GHz (AE), $f_0$=7.15 GHz (SE), $f_0$=5.78 GHz (AC), and $f_0$=6.44 GHz (SC) are obtained for $J_{DC}$=-1.6 x $10^7$ A/cm$^2$, the microwave central frequency $f_{AC0}$ is directly displayed in the figure. In Fig. 4(b) the oscillator frequencies are $f_0$=5.82 GHz (AE) for $J_{DC}$=-1.6 x $10^7$ A/cm$^2$, $f_0$=7.15 GHz (SE) for $J_{DC}$=-1.6 x $10^7$ A/cm$^2$, $f_0$=6.18 GHz (AC) for $J_{DC}$=-1.9 x $10^7$ A/cm$^2$, and $f_0$=5.82 GHz (SC) for $J_{DC}$=-1.3 x $10^7$ A/cm$^2$, in this case we plot different bias currents to show the better locking properties in the 2-harmonic. For both configuration of polarizers (symmetric and asymmetric), the circular structure has a locking region larger than the one in the ellipse. In addition, the locking bandwidth exhibits symmetrical response with respect to the central oscillator frequency characterized by a linear relationship as function of $J_M$, this result is in agreement with analytical formulation achieved in the approximation of "weak" microwave signal.[22] As $J_M$ increases, the synchronization regions are not symmetric as expected due to the non-linearities of the STO (see Fig. 4(a) AE and Fig. 4(b) SC).[25] From general point of view, the simpler explanation of the locking properties of STOs is based on the fact that the oscillator phase is only marginally stable with respect to the phase fluctuations and consequently also small perturbations can lead to a substantial change of the phase dynamics at long times.[22]

The injection locking in STOs for frequencies larger than the oscillator frequency[11] has been demonstrated both for vortex-dynamics[13,26] and for uniform modes excited in devices biased with an in-plane bias field.[27]

Differently, half-harmonic locking has not been achieved experimentally or predicted since the phenomenon of frequency modulation can occur up to 90% of the oscillator frequency.[28] In order to understand the half-locking mechanism, we studied a comparative test by performing micromagnetic simulations on a spin valve composed by a single free layer and one perpendicular polarizer. Our results indicate the presence of locking to the first and the second harmonic, but there is not half-harmonic synchronization. With this in mind, we can underline that the sub-harmonic locking behavior of the two free layers system is due to the torque from the perpendicular polarizer, responsible for the synchronization at the frequency of the mode excited in the free layer.

To identify the differences among different locking states and the free running oscillation data, we analyzed the time domain response including the time evolution of the spatial distribution of the magnetization. Fig. 5 shows the average $x$-component of the magnetization of the top free layer (AC structure) driven by $J_{DC}$=-1.9 x $10^7$ A/cm$^2$ without microwave signal (blue-dashed line), and for $J_M$=5.1 x $10^6$ A/cm$^2$ at different frequency $f_{AC0}$=3.05 GHz (black solid line), 6.15 GHz (red dash-dotted line), and 11.52 GHz (green dotted line) (same scenario of Fig. 3). As can be observed the time domain traces at regime have the same amplitude and frequency with an offset of the oscillation phase, with either delay (0.5:1 and 2:1 locking) and lead (1:1 locking) angle depending on the transient regime (not shown). Fig. 6 displays the snapshots (point A-F as indicated in Fig. 5) of the magnetization (the first and second line are for the top ($M_1$) and the bottom ($M_2$) layer respectively) for the AC structure ($J_{DC}$=-1.9 x $10^7$ A/cm$^2$, $J_M$= 5.1 x $10^6$ A/cm$^2$ and $f_{AC0}$=3.05 GHz). The same magnetization patterns are observed for the other time domain data by considering the right phase. In other words, for $J_M$=0 the magnetization configuration A is at $t=t_0$, in presence of the microwave source ($J_M$=5.1 x $10^6$ A/cm$^2$) the point A is at $t=(t_0-0.05)$ ns, $t=(t_0+0.07)$ ns, and $t=(t_0+0.19)$ ns for the locking 0.5:1, 1:1, and 2:1 respectively (see Fig. 5). The same consideration is valid for the configurations B-F. As can be noted, the excited mode is non-uniform and the two free-layers are phase-locked within the same scenario as explained in Ref.[9].

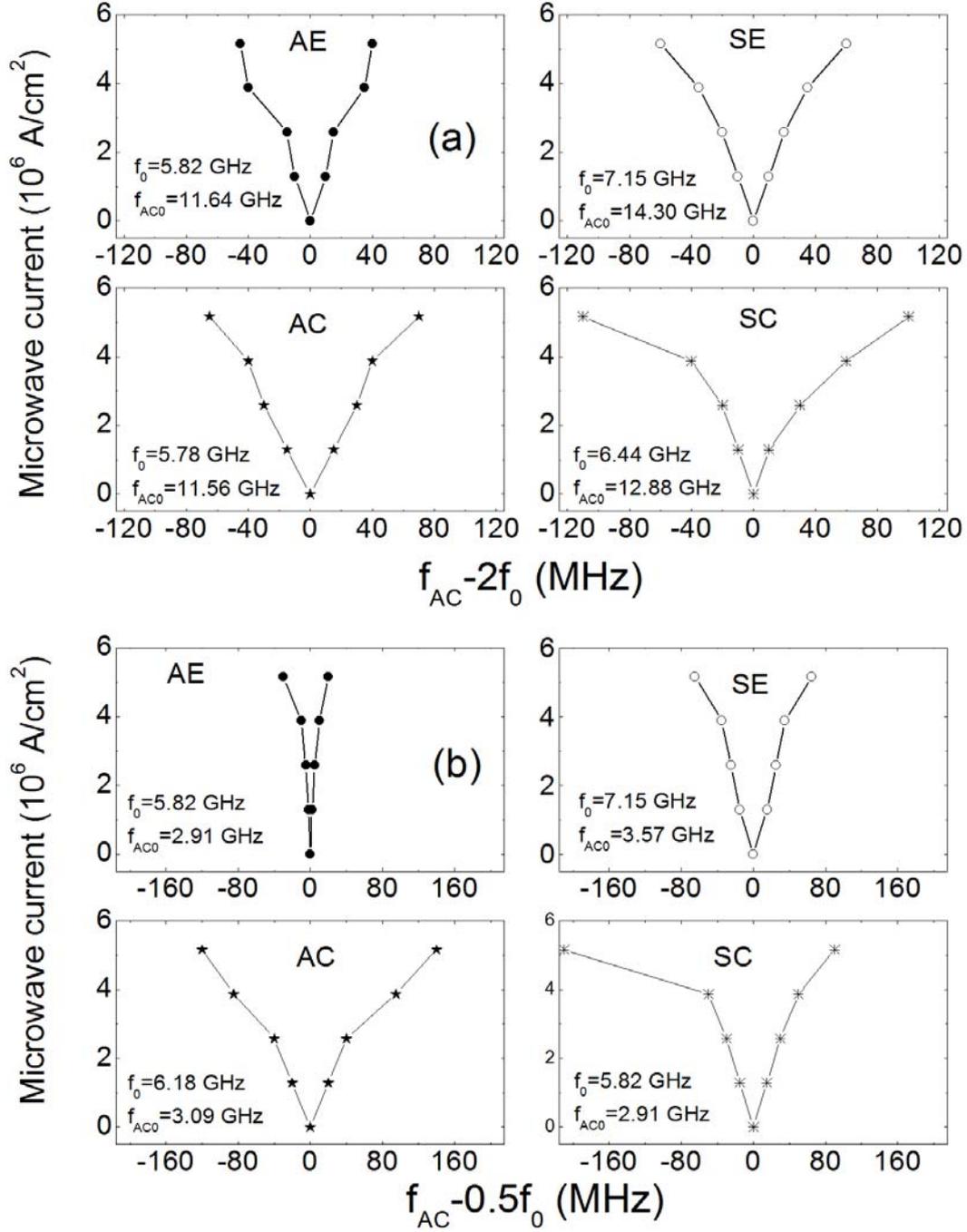

FIG. 4 (a) Arnold Tongue of the four structures at the second harmonic component at zero field and $J_{DC}$=-1.6 x $10^7$ A/cm$^2$. (b) Arnold Tongue of the four structures at half harmonic at zero field and different $J_{DC}$, AE: -1.6 x $10^7$ A/cm$^2$, SE: -1.6 x $10^7$ A/cm$^2$, AC: -1.9 x $10^7$ A/cm$^2$, SC: -1.3 x $10^7$ A/cm$^2$. $f_0$ is the oscillator frequency at that bias current and $f_{AC0}$ is the microwave central frequency.

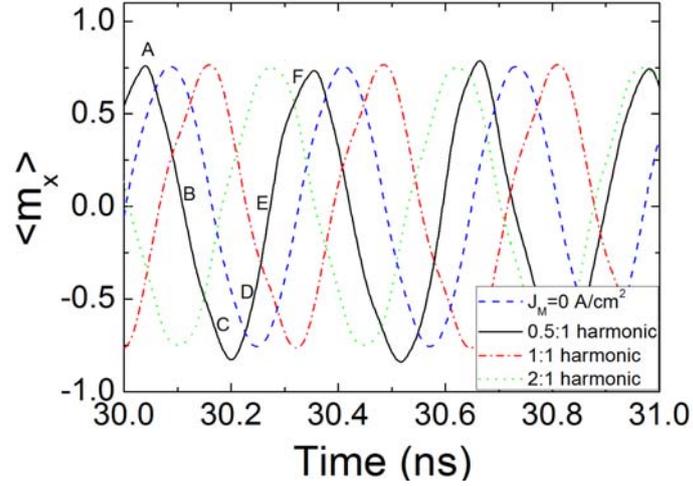

FIG. 5 Temporal evolution of the *x*-component of the magnetization of the top layer for the AC structure when $J_{DC}$=-1.9 x $10^7$ A/cm$^2$. The four different curves are for $J_M$=0 A/cm$^2$ (blue-dashed line),and for a fixed $J_M$=5.1 x $10^6$ A/cm$^2$ at $f_{AC0}$=3.05 GHz (black solid line), $f_{AC0}$=6.15 GHz (red dash-dotted line), and $f_{AC0}$=11.52 GHz (green dotted line).

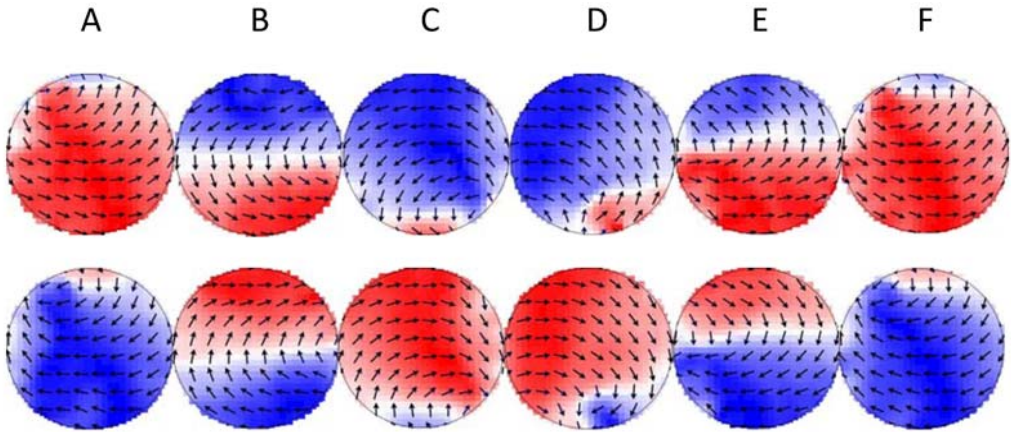

FIG. 6 Snapshots of the magnetization of the two free layers (first and second line for $M_1$ and $M_2$ respectively) for the AC structure when $J_{DC}$=-1.9 x $10^7$ A/cm$^2$ and the microwave source is $J_M$=

5.1 x $10^6$ A/cm$^2$ and $f_{AC0}$=3.05 GHz. The points A-F are related to the time traces indicated in Fig. 5. (the color red and blue are positive and negative-$x$ direction).

In summary, this work identifies a STO configuration where the synchronization to a microwave current is achieved with no bias field. In addition, the locking regions are attained for not only the oscillator frequency $f_0$, but also at 0.5$f_0$ and 2$f_0$. Those properties make this system promising for applicative point of view, in particular for the next generation of low power and small size on-chip devices such as high-speed (the locking transient is of the order of ns (not shown)) frequency dividers and multipliers, and phase-locked-loop demodulators[14]. In addition, the injection locking on those STOs can be used to achieve future clock generators with lower phase noise. Finally, our results indicate that the locking properties of the two free-layer spin-valves are universal and qualitatively independent on cross sectional area and perpendicular polarizer materials.

**ACKNOWLEDGMENTS**

This work was supported by project MAT2011-28532-C03-01 from Spanish government, project SA163A12 from Junta de Castilla y Leon, and project PRIN2010ECA8P3 from Italian MIUR.